\documentclass[amsmath,amssymb,pra,twocolumn]{revtex4}
\usepackage[cp1251]{inputenc}
\usepackage[english]{babel}
\usepackage{color}
\usepackage{graphicx}
\usepackage{dcolumn}
\usepackage{braket}
\usepackage{amsmath}

\usepackage{natbib}

\begin{document}
\bibliographystyle{unsrt}

\title{Noncollinear biphoton states: enormously high resource of azimuthal entanglement as it's seen in Cartesian variables.}

\author{M.V. Fedorov$^{1,2}$}
\email{fedorovmv@gmail.com}
\address{$^1$A.M.~Prokhorov General Physics Institute,
 Russian Academy of Sciences, 38 Vavilov st., Moscow, 119991, Russia}
 \address{$^2$National Research University Higher School of Economics, 20 Myasnitskaya Ulitsa,
Moscow, 101000, Russia}

\date{\today}

\begin{abstract}
Single-particle and coincidence distributions of photons are analyzed for the noncollinear frequency-degenerate type-I regime of Spontaneous Parametric Down-Conversion. Noncollinearity itself is shown to provide a new mechanism of strong broadening of the single-particle distributions in Cartesian components of the photon's transverse wave vectors. Related to this, the degree of entanglement appears to be very high and, in fact, this is the same enormous resource of azimuthal entanglement which was found to occur in the formalism of spherical angles used for characterization of photon wave vectors (Phys. Rev. A, {\bf 93}, 033830, 2016). In Cartesian variables this phenomenon manifests itself as a strong broadening and a very unusual and peculiar shape of the arising single-particle distribution curves. In theory, the key reason for these effects is the reduction of the total wave function of two photons over one of two orthogonal degrees of freedom. In the suggested and discussed experimental scheme this means that all photons of the emission cone have to be taken into account rather than only photons propagating in one given plane which is a common practice in many experiments.
\end{abstract}

\pacs{32.80.Rm, 32.60.+i}
\maketitle

\section{Introduction}

\section{Introduction}

A structure of emission in the type-I noncollinear Spontaneous Parametric Down-Conversion (SPDC) is well known: SPDC photons propagate along a cone with the axis ($0z$) coinciding with the central propagation direction of the pump \cite{Shih,Torres,Kim}. Quantum mechanically, all manifold of emitted SPDC photons is characterized by their wave function depending on transverse components of photon wave vectors ${\vec k}_{1\,\perp}\perp 0z$ and ${\vec k}_{2\,\perp}\perp 0z$, where the indices 1 and 2 indicate two indistinguishable SPDC photons. Each of two SPDC photons has two degrees of freedom, for example, corresponding to motions in $0x$ and $0y$ directions. In this specific case the biphoton wave function depends on two pairs of variables, $k_{1,2\,x}$ and $k_{1,2\,y}$. Alternatively, at given values of the total wave vectors vectors $k_{1,2}$, their transverse components ${\vec k}_{1,2\,\perp}$ can be characterized by the spherical angles of the total wave vectors ${\vec k}_i$,  by polar (zenith) angles $\theta_{1,2}$ defined as angles between ${\vec k}_{1,2}$  and the $z$-axis, and azimuthal angles $\alpha_{1,2}$ defined as angles between ${\vec k}_{1,2\,\perp}$ and the $x$-axis. This spherical-angle parametrization was used in the recent work \cite{PRA-16}, where the azimuthal entanglement (in variables $\alpha_{1,2}$) was found to be extremely high at sufficiently high degree of noncollinearity. The main physical reason for this conclusion is in the occurring in this case axial symmetry of emission of SPDC photons. A scheme for practical use of this effect was suggested. On the other hand, as the degree of entanglement has to be invariant with respect to change of a basis used for analysis \cite{Miatto}, it's rather interesting to find out how the extremely high azimuthal entanglement shows itself in Cartesian variables. This is the topic of this work. It will be shown that the enormous azimuthal entanglement manifests itself in the Cartesian-variable picture in the form of a very strong broadening of the single-particle distribution in components of transverse wave vectors, and the corresponding distribution curves will be shown to have a rather unusual and peculiar shape. A scheme for observing these effects  experimentally will be discussed.

\section{Biphoton wave function in Cartesian variables}

Thus, let us consider the type-I frequency-degenerate SPDC process in which a cw-laser pump with a wavelength $\lambda_p$ propagates in a crystal as a vertically polarized extraordinary wave whereas emitted photons are horizontally polarized, propagate in a crystal as ordinary waves, and have wavelengths $2\lambda_p$.  For such processes a general expression for the transverse-momentum biphoton wave function is well known \cite{PRA-16,Monk,Law-E,2007,2008}
\begin{equation}
  \Psi\propto E_p({\vec k}_{p\,\perp})\,{\rm sinc}\left(\frac{L\Delta}{2}\right),
 \label{Psi}
\end{equation}
where ${\rm sinc}(x)=\sin x/x$, $L$ is the crystal length, $E_p({\vec k}_{p\,\perp})$ is the Fourier transform of the pump transversal amplitude $E_0({\vec r}_\perp)$. In the approximation of a sufficiently wide crystal ${\vec k}_{p\,\perp}={\vec k}_{1\,\perp}+{\vec k}_{2\,\perp}$, and in the case of a Gaussian pump envelope with the waist $w_p$
\begin{equation}
 \label{Gaussian}
  E_p\propto\exp{\left(-\frac{w_p^2({\vec k}_{1\,\perp}+{\vec k}_{2\,\perp})^2}{2}\right)}.
\end{equation}
Note that for shortening notations we drop as unimportant for further derivations all constant coefficients in Eqs.(\ref{Psi}), (\ref{Gaussian}) and other equations below by keeping only parts depending on the variables ${\vec k}_{1\,\perp}$, ${\vec k}_{2\,\perp}$.

The second factor on the right-hand side of Eq. (\ref{Psi}) is related to formation of SPDC emission in a crystal, and in this term $\Delta$ is the phase mismatch. In the paraxial approximation
\begin{equation}
 \label{Delta}
 \Delta=k_{p\,z}-k_{1\,z}-k_{2\,z}\approx \Delta_0+ \frac{\left({\vec k}_{1\perp}-{\vec k}_{2\perp}\right)^2}{4k_1}
\end{equation}
with $\Delta_0$ being the zero-order term in the expansion of $\Delta$ in powers of ${\vec k}_{1,2\,\perp}$
\begin{equation}
 \label{Delta-0}
 \Delta_0=k_p-k_1-k_2=\frac{2\pi}{\lambda_p}\,(n_p-n_o)\equiv\frac{2\pi}{\lambda_p}\Delta n
\end{equation}
where $n_p$ and $n_o$ are the refractive indices of the pump and of emitted photons at the wave lengths,correspondingly, $\lambda_p$ and $2\lambda_p$, and $\Delta n=n_p-n_o$. If the ordinary-wave refractive index is isotropic, the refractive index of the pump varies with varying orientation of a crystal and direction of the pump wave vector ${\vec k}_p$. In a general case \cite{LL}
\begin{equation}
 \label{np}
 n_p(\lambda_p,\vartheta_p)=\frac{n_o(\lambda_p)n_e(\lambda_p)}
 {[n_o^2(\lambda_p)\sin^2\vartheta+n_e^2(\lambda_p)\cos^2\vartheta]^{1/2}},
\end{equation}
where $\vartheta$ is the angle between the pump wave vector ${\vec k}_p$ in a crystal and its optical axis ($OA$), and $n_e(\lambda_p)$ is the extraordinary-wave refractive index for the direction perpendicular to $OA$ ($\vartheta=\pi/2$). For direction along $OA$, $\vartheta=0$ and $n_p=n_o(\lambda_p)$. Explicit expressions for the functions $n_o(\lambda)$ and $n_e(\lambda)$ are given by Sellmeier formulas \cite{Dmitr}, from which one easily finds that, e.g., in  a BBO crystal at the pump wave length  $\lambda_p=0.4047\mu{\rm m}$, the constants $n_{e,o}(\lambda_p)$ and $n_o(2\lambda_p)$ are equal to  $n_e(\lambda_p)=1.56801$ $n_e(\lambda_p)=1.56801$, $n_o(\lambda_p)=1.69236$, and $n_o(2\lambda_p)=1.66109$.

\begin{figure}[h]
\centering\includegraphics[width=7cm]{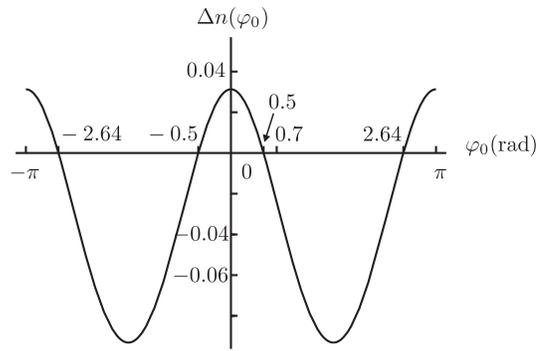}
\caption{{\protect\footnotesize {The difference of the refractive indices $\Delta n(\varphi_0)$ for a BBO crystal and the pump wavelength $\lambda_p=0.4047\,\mu{\rm m}$.}}}\label{Fig1}
\end{figure}

In a general case, the angle $\vartheta$ between the pump wave vector ${\vec k}_p$ and optical axis is determined by both the angle $\varphi_0$ with $OA$ and the $z$-axis  and by direction of the pump wave vector ${\vec k}_p$. But, as always $w\gg \lambda_p$, directions of the pump wave vector ${\vec k}_p$ are always close to the central propagation axis of the pump $0z$. If, besides, the degree of noncollinearity is sufficiently well pronounced, in accordance with estimates of Ref. \cite{PRA-16}, deviations of ${\vec k}_p$ from the $z$-axis are not important and can be ignored. This is the so called "No Walk-Off" (NWO) approximation of Ref. \cite{PRA-16}, which is reasonably good if the cone opening angle $\theta_0$ is not too small. At ${\vec k}_p\| 0z$ the angle $\vartheta$ in Eq. (\ref{np}) coincides with the angle $\varphi_0$ between $OA$ and the $z$-axis, and both the refractive index $n_p$ and the difference of refractive indices $\Delta n$ can be considered as functions of only one angle $\varphi_0$. For  BBO crystal at the pump wave length $\lambda_p=0,40407\,\mu{\rm m}$ the function $\Delta n(\varphi_0)$ is plotted in Fig. \ref{Fig1}.

The case   $\Delta_0=\Delta n =0$ ($n_p=n_0$) corresponds to the collinear regime of SPDC, which occurs for this crystal and the chosen pump wavelength at $\varphi_{0\,{\rm Coll}}\approx 0.5\, {\rm rad}$. The noncollinear regime of SPDC occurs at $n_p < n_0$, or at $\varphi_{0\,{\rm Coll}}<|\varphi_0|<\pi-\varphi_{0\,{\rm Coll}}$. The angles  $\varphi_0=0.1\,{\rm and}\,0.7\,{\rm rad}$ shown in Fig. \ref{Fig1} correspond to two specific examples considered below.

\begin{figure}[h]
\centering\includegraphics[width=7.5cm]{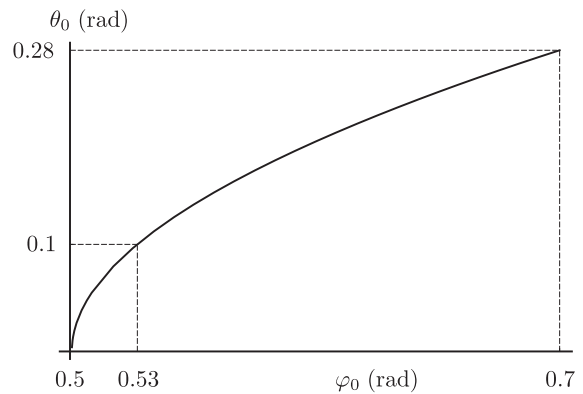}
\caption{{\protect\footnotesize {The cone opening angle $\theta_0$ as a function of the angle $\varphi_0$ between the optical axis of a crystal and the $z$-axis.}}}\label{Fig2}
\end{figure}

In the noncollinear regime, the difference of refractive indices determines the opening angle of the emission cone of SPDC \cite{PRA-16}:
\begin{equation}
 \label{opening angle}
 \theta_0=\sqrt{2n_o(n_o-n_p)}\equiv \sqrt{-2n_o\Delta n}.
\end{equation}
As $n_p$ and $\Delta n$ in this definition depend on the angle $\varphi_0$ between $OA$ and the central propagation axis of the pump $0z$, the cone opening angle $\theta_0$ varies also with variation of $\varphi_0$, and the function $\theta_0(\varphi_0)$ can be found to have the form presented in Fig.\ref{Fig2}

\noindent Also from the definition of the opening angle (\ref{opening angle}) one can find a simple interpolation formula fitting pretty well the curve in Fig. \ref{Fig2}
\begin{equation}
 \label{interpolation}
 \theta_0(\varphi_0)=0.63\sqrt{\varphi_0-0.5008}.
\end{equation}
Two above-mentioned values of $\varphi_0$, $0.5275\approx 0.53$ and $0.7$, correspond to  $\theta_0=0.1\,{\rm and}\, 0.28\,{\rm rad}$.

The squared sum and difference of transverse wave vectors of photons in Eqs. (\ref{Psi}) and (\ref{Delta}) can be expressed in terms of sums and differences of their $x$- and $y$-components
\begin{gather}
  \nonumber
 \left({\vec k}_{1\perp}\pm{\vec k}_{2\perp}\right)^2=\left(k_{1\,x}\pm k_{2\,x}\right)^2+\left(k_{1\,y}\pm k_{2\,y}\right)^2\\
 \equiv k_{\pm\,x}^2+k_{\pm\,y}^2,
 \label{sums-diffs}
\end{gather}
where $k_{\pm\,x}=k_{1\,x}\pm k_{2\,x}$ and $k_{\pm\,y}=k_{1\,y}\pm k_{2\,y}$.

Note that owing to validity of the NWO approximation at relatively large values of $\theta_0$, the SPDC emission cone is axial symmetric \cite{PRA-16} and, owing to this, a choice of two orthogonal axes $0x$ and $0y$ is arbitrary and not necessarily related to the horizontal and vertical directions.

With Eqs. (\ref{Psi})-(\ref{sums-diffs}) combined together, the general expression of the biphoton wave function takes the form
\begin{gather}
\nonumber
\Psi(k_{1\,x},k_{2\,x},k_{1\,y},k_{2\,y})\propto\\
\nonumber
\exp\left\{-\left[(k_{1\,x}+k_{2\,x})^2+(k_{1\,y}+k_{2\,y})^2\right]w^2/2\right\}\times\\
\label{Psi-gen}
{\rm sinc}\left[\frac{\pi L}{8n_o\lambda_p}\left(4\theta_0^2-\frac{\lambda_p^2}{\pi^2}
\left[(k_{1\,x}-k_{2\,x})^2+(k_{1\,y}-k_{2\,y})^2\right]\right)\right]
\end{gather}

\section{Biphoton probability densities.}

The squared absolute value of the wave function is the 4-dimensional bipartite probability density distribution
\begin{equation}
 \label{4D-distr}
 \frac{dW}{dk_{1\,x}dk_{2\,x}dk_{1\,y}dk_{2\,y}}=|\Psi(k_{1\,x},k_{2\,x},k_{1\,y},k_{2\,y})|^2.
\end{equation}
Four variables in this probability density correspond to two photons (1 and 2) each of which has two degrees of freedom ($x$ and $y$). Let us define the bipartite distribution in one of these two degrees of freedom ($x$) as the integral of the probability density (\ref{4D-distr}) over variables corresponding to the second degree of freedom ($y$).
\begin{equation}
 \label{2D-integrated}
 \frac{dW_{\rm red}}{dk_{1\,x}dk_{2\,x}}=\int dk_{1\,y}dk_{2\,y}|\Psi(k_{1\,x},k_{2\,x},k_{1\,y},k_{2\,y})|^2.
\end{equation}
This reduced probability density coincides exactly with the diagonal part of the reduced density matrix if reduction is understood as reduction with respect to one of two degrees of freedom of both photons
\begin{gather}
  \nonumber
 \rho_r(k_{1\,x},k_{2\,x};k^\prime_{1\,x},k^\prime_{2\,x})=\int dk_{1\,y}dk_{2\,y}\Psi(k_{1\,x},k_{2\,x},k_{1\,y},k_{2\,y})\\
 \label{rho-r}
 \times\Psi^*(k^\prime_{1\,x},k^\prime_{2\,x},k_{1\,y},k_{2\,y}),
\end{gather}
and
\begin{equation}
 \label{W-red-rho-r}
 \frac{dW_{\rm red}}{dk_{1\,x}dk_{2\,x}}\equiv
 \rho_r(k_{1\,x},k_{2\,x};k_{1\,x},k_{2\,x}).
\end{equation}
It should be stressed that though the definition of reduction with respect to one of two degrees of freedom of both photons is similar to the usual reduction over one of two subsystems or over variables of one of two particles, the difference is rather well pronounced. In principle, the procedure of reduction in one of several degrees of freedom can be applied even for single-particle states. For example, such situation occurs in the case of coherently excited superpositions of one-electron excited atomic states, which can be reduced over angular variables to give mixed radial one-electron states. This problem was considered in the work \cite{QE} where in this way the degree-of-freedom entanglement of one-electron atomic states was defined and evaluated.

Note also that the dependencies on the $x$- and $y$-components of wave vectors in the sinc-function in Eq. (\ref{Psi-gen}) do not factorize and, for this reason, integration over $k_{1,2\,y}$ is not equivalent at all to a simple substitution $k_{1,2\,y}=0$ in the general expression for the wave function ,
$$
dW_{{\rm red}}/dk_{1\,x}dk_{2\,x}\neq|\Psi(k_{1\,x},k_{2\,x},0,0|^2.
$$

Integration over $k_{1\,y}$ and $k_{2\,y}$ in Eq. (\ref{2D-integrated}) can be substituted by integration over $k_{-\,y}$ and $k_{+\,y}$ to give
\begin{equation}
  \frac{dW_{\rm red}}{dk_{1\,x}dk_{2\,x}}\propto e^{-w_p^2k_{+\,x}^2}F(k_{-\,x}),
 \label{Pro-dens via F}
\end{equation}
where
\begin{gather}
 \label{F-k min x}
 F(k_{-\,x})= \int dq\,{\rm sinc}^2\left[\frac{\pi L}{8n_o\lambda_p}\left(4\theta_0^2-\frac{\lambda_p^2k_{-\,x}^2}{\pi^2}-q^2\right)\right],\;\;
\end{gather}
and $q=\lambda_pk_{-\,y}/\pi$ is the dimensionless integration variable used instead of $k_{-\,y}$.

\begin{figure}[h]
\centering\includegraphics[width=8.5cm]{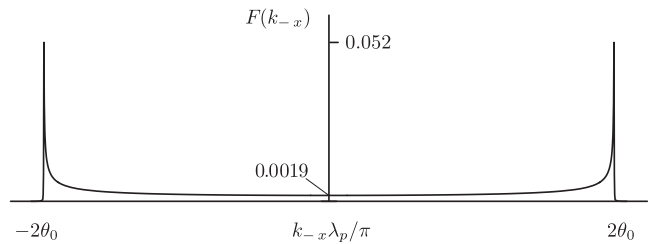}
\caption{{\protect\footnotesize {The function $F(k_{-\,x})$ (\ref{F-k min x}) in dependence on the dimensionless variable $\lambda_p k_{-\,x}/\pi$ at  $\theta_0=0.28\,{\rm rad}$ and $L=0.5\,{\rm cm}$.}}}\label{Fig3}
\end{figure}

In principle, integration in Eq. (\ref{F-k min x}) is doable analytically to give some expressions in terms of hypergeometric functions, which are too cumbersome and practically useless to be shown explicitly. Instead, the result of such calculation is shown in Fig. \ref{Fig3} for the same values of parameters which were used in Ref. \cite{PRA-16} and already mentioned above: BBO, $\lambda_p=0.4047\,\mu{\rm m}$, $\varphi_0=0.7\,{\rm rad}$, $\theta_0=0.28\, {\rm rad}$ and $L=0.5\,{\rm cm}$.

\noindent On the other hand, as the argument of the sinc$^2$-function in Eq. (\ref{F-k min x}) contains a very large parameter in front of round brackets, $L/\lambda_p\sim 10^4$,  the sinc$^2$-function can be approximated roughly by the delta-function to give
\begin{figure}[t]
\centering\includegraphics[width=8.5cm]{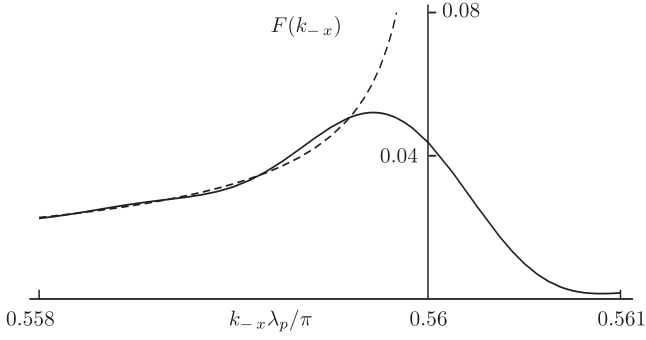}
\caption{{\protect\footnotesize {Exact (solid line, Eq. (\ref{F-k min x})) and approximate (dashed line, Eq. (\ref{F-appr})) functions $F(k_{-\,x})$ in a small vicinity of the point $2\theta_0$}.}}\label{Fig4}
\end{figure}
\begin{equation}
 \label{F-appr}
F_{\rm appr}(k_{-\,x})\approx \frac{8n_o\lambda_p}{L\sqrt{4\theta_0^2-\lambda_p^2k_{-\,x}^2/\pi^2}},
\end{equation}
The curves $F(k_{-\,x})$ determined by Eqs. (\ref{F-k min x}) (exact)  and (\ref{F-appr}) (approximate) are practically indistinguishable everywhere except very small vicinities of the points $\lambda_pk_{-\,x}/\pi=\pm 2\theta_0$. In Fig. \ref{Fig4} these two curves are plotted together in  a small vicinity of the point $\lambda_p k_{-\,x}/\pi=2\theta_0$. Any differences between the curves disappear already at $2\theta_0-\lambda_p k_{-\,x}/\pi\geq 0.002$, which is very close to $2\theta_0=0.56$. Besides, though the expression of Eq. (\ref{F-appr}) has singularities at $\lambda_p k_{-\,x}/\pi=\pm 2\theta_0$, these singularities are integrable, owing to which the approximate function $F_{\rm appr}(k_{-\,x})$ can be used successfully for calculation of any average values, contributions to which are given mainly by the whole variation interval $-2\theta_0\leq\lambda_p k_{-\,x}/\pi\leq 2\theta_0$ rather than by vicinities of the singular points $\pm 2\theta_0$.

As a whole, the curve of Fig. \ref{Fig3} is rather unusual: it has two very narrow and not too high peaks and very wide slowly varying plateau-type part in a wide region from  $-2\theta_0$ to $2\theta_0$. Its width equals roughly $4\theta_0$ or, in terms of the difference of momenta $k_{-\,x}$, $4\pi\theta_0/\lambda_p$. More rigorously the width of this distribution $\Delta k_{-\,x}$ can be estimated  as the square root of the average squared difference of momenta $k_{-\,x}=k_{1\,x}-k_{2\,x}$. This average can be calculated with the help of the weighting function $F_{\rm appr}(k_{-\,x})$ of Eq. (\ref{F-appr}) to give
\begin{equation}
 \label{k-sqared-av}
 \overline{k_{-\,x}^2}=\frac{\int dk_{-\,x}\, k_{-\,x}^2F_{\rm appr}(k_{-\,x})}{\int dk_{-\,x}\,F_{\rm appr}(k_{-\,x})}=\frac{2\pi^2\theta_0^2}{\lambda_p^2}
\end{equation}
and
\begin{equation}
 \label{width of F}
 \Delta k_{-\,x}=\sqrt{\overline{k_{-\,x}^2}}=\frac{\sqrt{2}\,\pi\theta_0}{\lambda_p}.
\end{equation}
This width is really huge. At $\theta_0=0.28\,{\rm rad}$ and $\lambda_p=0.4047\,\mu{\rm m}$ Eq. (\ref{width of F}) yields
$\Delta k_{-\,x}=30739\,{\rm cm}^{-1}$.

\section{single-particle and coincidence distributions and the degree of entanglement}

Thus, the bipartite distribution in the transverse components of the wave vectors $k_{1\,x}$ and $k_{2\,x}$ is given by Eq. (\ref{Pro-dens via F}) which can be rewritten as
\begin{equation}
 \label{Bipart dist}
  \frac{dW_{\rm red}}{dk_{1\,x}dk_{2\,x}}\propto e^{-w_p^2(k_{1\,x}+k_{2\,x})^2}F(k_{1\,x}-k_{2\,x}).
\end{equation}
At a given value of, e.g.,  $k_{2\,x}$ Eq. (\ref{Bipart dist}) determines the conditional probability density of seeing a photon with a varying  $x$-projection of the wave vector $k_{1\,x}$
\begin{gather}
  \nonumber
  \frac{dW^{(c)}(k_{1\,x})}{dk_{1\,x}}=
  \left.\frac{dW_{\rm red}}{dk_{1\,x}dk_{2\,x}}\right|_{k_{2\,x}={\rm const}.}\propto e^{-w_p^2(k_{1\,x}+k_{2\,x})^2} \\
  \label{Coinc}
  \times F(k_{1\,x}-k_{2\,x})\approx e^{-w_p^2(k_{1\,x}+k_{2\,x})^2}\times F(2k_{2\,x}).
\end{gather}
In dependence on $k_{1\,x}$, this expression determines  a narrow curve centered at $k_{1\,x}=-k_{2\,x}$ and having the width
\begin{equation}
 \label{width-Gauss}
 \Delta k_{1\,x}^{(c)}=\sqrt{\overline{k_{+\,x}^2}}=\frac{1}{2w_p},
\end{equation}
where averaging is taken with the Gaussian function $e^{-w_p^2 k_{+\,x}^2}$. Typically, the width  $\Delta k_{1\,x}^{(c)}$
is very small compared to $\Delta k_{-\,x}$ (\ref{width of F}) if only the SPDC noncollinearity is sufficiently well pronounced. For example, at $w_p=0.5\,{\rm cm}$ Eq. (\ref{width-Gauss}) gives $\Delta k_{1\,x}^{(c)}=1\,{\rm cm}^{-1}$, which is more than 4 orders of magnitude smaller than $\Delta k_{-\,x}\sim 3\times 10^4\,{\rm cm}^{-1}$ (\ref{width of F}).

For getting a single-particle distribution of photons in $k_{x\,1}$ from the bipartite probability density (\ref{Bipart dist}) one has to integrate the latter over $k_{x\,2}$. As the function $F(k_{-\,x})$ in Eq. (\ref{Pro-dens via F}) is much wider than the Gaussian function $e^{-w_p^2k_{+\,x}^2}$, this last exponent can be approximated by the $\delta$-function, $\delta(k_{1\,x}+k_{2\,x})$, to give
\begin{equation}
 \label{Dist-single}
 \frac{dW^{(s)}(k_{1\,x})}{dk_{1\,x}}=F(2k_{1\,x}).
\end{equation}
The width of this distribution is
\begin{equation}
 \label{single-width}
 \Delta k_{1\,x}^{(s)}=\frac{\Delta k_{-\,x}}{2}= \frac{\pi\theta_0}{\sqrt{2}\,\lambda_p}.
\end{equation}
As known \cite{2004,2006}, the ratio of the single-particle to coincidence widths is a reasonably good entanglement quantifier for states with continuous variables , and in the case under consideration this parameter is given by
\begin{equation}
 \label{R}
 R=\frac{\Delta k_{1\,x}^{(s)}}{\Delta k_{1\,x}^{(c)}}=\frac{\pi\sqrt{2}\,\theta_0\,w_p}{\lambda_p}\gg 1.
\end{equation}
Numerically, at the same values of parameters which were used in estimates of the widths (\ref{width of F}) and (\ref{width-Gauss}), Eq. (\ref{R}) gives $R\approx 1.5\times 10^4$. This is the same result that has been obtained for azimuthal entanglement in the frame of analysis in the basis of spherical angles \cite{PRA-16}. Hence, both considerations, in Cartesian and in spherical-angle variables describe the same phenomenon of extremely high azimuthal entanglement occurring owing to the axial symmetry of emission in the noncollinear type-I SPDC process.

As seen from Eq. (\ref{R}), for the described type of entanglement the degree of entanglement is controlled completely by only two factors: by the opening angle $\theta_0$ of the emission cone and by the pump waist $w$, and it does not depend at all of the crystal length $L$. This differs essentially noncollinear SPDC process from the collinear regime, where the degree of entanglement is determined by competition of the pump waist and crystal length \cite{Law-E}. In the noncollinear regime, noncollinearity of the process itself provides a much greater broadening of the momentum single-particle distribution which exceeds significantly the broadening effect related to the finite length of a crystal and makes the entanglement parameter $R$ independent of $L$.

\section{Observability in experiments with transverse-momentum distributions}
\begin{figure}[h]
\centering\includegraphics[width=8cm]{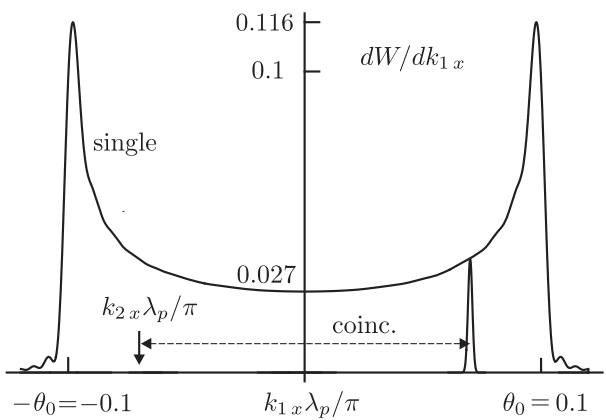}
\caption{{\protect\footnotesize {Coincidence and single-particle distributions in $k_{1\,x}$ ( in units of $\pi/\lambda_p$) for the pump and crystal parameters $w=L=10^{-1}\,{\rm cm}$, $\varphi_0=0.5275{\,\rm rad}$, and $\theta_0=0.1{\rm rad}$.}}}\label{Fig5}
\end{figure}

In the following analysis of a very high entanglement in noncollinear SPDC, the set of pump and crystal parameters will be chosen not as extreme as in the previous case and, probably, easier realizable experimentally: the pump waist $w=10^{-1}{\rm cm}$, the cone opening angle $\theta_0=0.1{\rm rad}\approx 5.7^\circ$ which corresponds to the angle between $OA$ and the central pump-propagation direction $0z$ equal to $\varphi_0=0.5275{\rm rad}$, and the crystal length $L=10^{-1}\,{\rm cm}$. The picture of the single-particle and coincidence distributions arising in this case is shown in Fig. \ref{Fig5}.

\begin{figure}[t]
\centering\includegraphics[width=7cm]{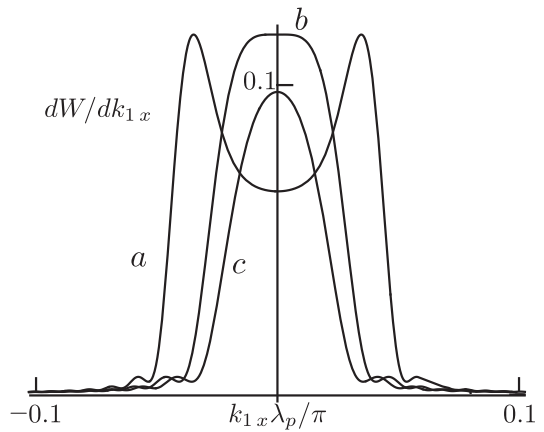}
\caption{{\protect\footnotesize {Single-particle distribution in $k_{1\,x}$ in the cases $ \theta_0=0.04\,(a)$; $\theta_0=0.02\,(b)$ and $\theta_0=0\,(c)$ .}}}\label{Fig6}
\end{figure}

\noindent At the chosen values of all parameters, the widths of the single-particle and conditional (coincidence) distributions (\ref{single-width}) and (\ref{width-Gauss}) are equal to
\begin{equation}
 \label{widths-moderate}
 \Delta k_{1\,x}^{(s)}=5489\,{\rm cm}^{-1}\quad{\rm and}\quad\Delta k_{1\,x}^{(c)}=5\,{\rm cm}^{-1}.
\end{equation}
The ratio of these widths is the parameter $R$ (\ref{R}) characterizing the degree of entanglement, and now it is not as enormously high as in the pervious estimates but, still, very high,
\begin{equation}
 \label{R-110}
 R=\frac{\Delta k_{1\,x}^{(s)}}{\Delta k_{1\,x}^{(c)}}\approx 1099\gg 1.
\end{equation}
In terms of Cartesian components of wave vectors this high entanglement is related to a very strong broadening of the single-particle distribution curve $dW/dk_{1\, x}$. This broadening and the very unusual shape of the distribution curve in Fig. \ref{Fig5} arise owing to a sufficiently pronounced noncollinearity of the SPDC process. In the case of sufficiently large opening angle of the SPDC cone $\theta_0$, the noncollinearity-induced broadening appears to be independent of the length of a crystal $L$ and significantly exceeds the width of the same curve in the collinear regime $\Delta (\lambda_p k_{1\,x}^{{\rm coll}})\sim\sqrt{\lambda_p/L}$. Variations of the single-particle distribution curves $dW(k_{1\, x})/dk_{1\, x}$ with decreasing values of the cone-opening angle $\theta_0$ is illustrated by three curves in the picture of Fig. \ref{Fig6}.
The curve $a$ corresponds to the case $\theta_0=0.04\approx 2.3^\circ$. This curve is similar to that of Fig. \ref{Fig5} though now two peaks of the curve are closer to each other and the dip between the peaks is not as deep as  in the case $\theta_0=0.1$. But still, the curve $a$ corresponds to the noncollinearity-induced regime of broadening the single-particle distribution. The curve $b$ corresponds to $\theta_0=0.02\approx 1.15^\circ$, and it describes the broadening regime intermediate between those related to noncollinearity and to a finite length of a crystal. In this case the peaks of the curve at $\pm\theta_0$ are so close to each other and the dip between them is so shallow that together they provide only the flat-top structure of the curve. At last, the curve $c$ in Fig. \ref{Fig6} corresponds to the purely collinear regime, $\theta_0=0$, and the width of this curve is estimated qualitatively as $\Delta (\lambda_pk_{1\,x}^{{\rm coll}})\sim \sqrt{\lambda_p/L}$. This analysis permits to formulate the following applicability condition of the described regime of strong noncollinearity-induced broadening of the single-particle distribution curves
\begin{equation}
 \label{condition}
 \theta_0\gg\sqrt{\frac{\lambda_p}{L}}.
\end{equation}

\begin{figure}[h]
\centering\includegraphics[width=6cm]{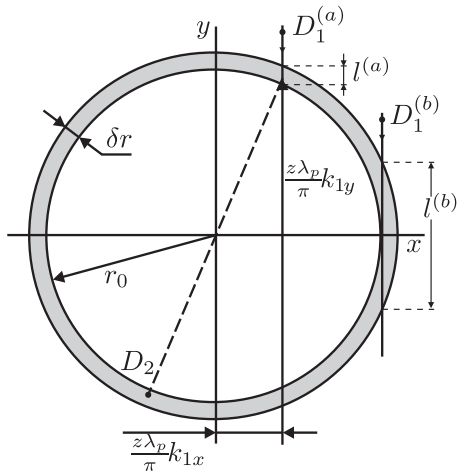}
\caption{{\protect\footnotesize {Section of the emission cone by a plane $(xy)\perp 0z$; $r_0 $ and $\delta r$ are the radius and thickness of the ring, and the dashed line is its diameter; $D_1^{(a,b)}$ and $D_2$  are detectors; vertical lines indicate directions of scanning the detectors $D_1^{(a,b)}$; position of vertical lines is shown to be parametrized by varying values of $k_{1\,x}$; $l^{(a,b)}$ are lengths ways at which pohotons are registered and summed during scanning.}}}\label{Fig7}
\end{figure}

Some further details of the assumed measurements can be explained most clearly in the following consideration of detector location and scanning in the section of the cone by a plane perpendicular to the $z$-axis (Fig. \ref{Fig7}).
This section is a ring with the radius $r_0$ and thickness $\delta r$ given by
\begin{equation}
 \label{ring}
 r_0=z\theta_0=10\,{\rm cm},\, \delta r=z \frac{\Delta k_{1\,x}^{(c)}\lambda_p}{\pi}=0.016\,{\rm cm},
\end{equation}
where the crystal-detector distance $z$ is taken equal $1\,{\rm m}$ and all other parameters are are the same as in Fig. \ref{Fig5}. $D_1$ and $D_2$ are detectors with apertures smaller than $\delta r$. For measuring single-particle distribution, the detector $D_2$ is not needed. Then, the detector $D_1$ has to move slowly along one of vertical lines ($\|0y$) counting all met photons and summing all counts. Positions of vertical lines  for a moving detector $D_1$ are characterized by values of the $x$- components of the transverse wave vector $k_{1\,x}$, as shown in Fig. \ref{Fig7} for the detector $D_1^{(a)}$. The sum of all counts at a given $k_{1\,x}$ determines one point at the curve $dW^{(s)}(k_{1\,x})/dk_{1\,x}$ of Fig. \ref{Fig5}. Then the position of the vertical line has to be changed with all the procedure repeated to get another point at the same curve, etc., etc.

Summation of counts at different positions of $D_1$ at a given vertical line corresponds to realization in experiment of the reduction procedure over the  $y$-degree of freedom in theory [Eq. (\ref{2D-integrated})].  The total amount of counts at any given $k_{1\,x}$ will be proportional to the length of a way $l(k_{1\,x})$ crossed by the detector through the thickness of the ring. Clearly enough, $l(k_{1\,x})$ is minimal and equals $\delta r$ at $k_{1\,x}=0$ when the vertical line for scanning the detector $D_1$ coincides with the $y$-axis. Oppositely, $l(k_{1\,x})$ is maximal at maximal values of $k_{1\,x}$, $k_{1\,x}^{({\rm max})}\approx\theta_0\pi/\lambda_p $, as shown in Fig. \ref{Fig7} for the detector at the line $D_1^{(b)}$. This increase of the length of a way $l(k_{1\,x})$ at $|k_{1\,x}|$ approaching its maximal values explains appearance of picks at the curves $dW^{(s)}/dk_{1\,x}$ in Figs. $3,\,5,\,6$.

If the detector $D_2$ is turned on at some position on the ring and if signals from the detectors $D_2$ and $D_1^{(a)}$ are sent to the coincidence scheme, a similar procedure can be used for measuring a very narrow coincidence curve $dW^{(c)}(k_{1\,x})/dk_{1\,x}$ by means of scanning the detector $D_1^{(a)}$ along a vertical line, summing all counts and repeating all this at a series of different very closely located vertical lines. The coincidence distribution will be obtained as the plot of total amounts of counts in dependence on positions of vertical lines parametrized by $k_{1\,x}$. The peak of the curve is expected to occur when the vertical line for the scanning the detector $D_1^{(a)}$ in Fig. \ref{Fig7} crosses the point symmetric to the position of the detector $D_2$, when both detectors are located at the opposite ends of the ring-diameter shown in the picture by the dashed line.

Note that instead of using detectors with aperture smaller than $\delta r$ one can use a wider-aperture detector (larger than $\delta r$ but, still, smaller than $r_0$) combined with installed in front of it narrow vertically oriented slit with the horizontal size smaller than $\delta r$. In this case one does not have to move detectors in the vertical direction and to sum counts, because automatically all photons in a slit will be counted and summed together. Such measurements have to be repeated at different horizontal positions of the slit+detector device to reproduce the single- and coincidence distribution curves.

\begin{figure}[h]
\centering\includegraphics[width=6cm]{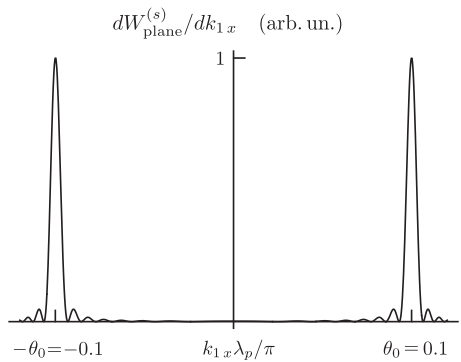}
\caption{{\protect\footnotesize {Expected single-particle distribution in the case registration only photons propagating in the plane $(x,z)$, with all other photons ignored.}}}\label{Fig8}
\end{figure}

Apart from further technical details, the most important feature of the discussed scheme of measurements, as well as of the theoretical analysis<, consists in contributions of all photons in the emission cone taken into account. This contrasts with the usual approach when only photons propagating in a given plane are taken into account (either $(xz)$ or $(yz)$ as, e.g., in Ref. \cite{2008}). In the theoretical description such simplification would correspond to exclusion of the reduction of the probability density over one of two degrees of freedom and to replacement of $dW_{\rm red}/dk_{1\,x}dk_{2\,x}$ (\ref{Pro-dens via F}) by $|\Psi(k_{1\,x},k_{2\,x},0,0)|^2$. This would give the single-particle distribution of the form
\begin{equation}
 \label{plane}
 \frac{dW_{\rm plane}^{(s)}}{dk_{1\,x}}\propto \int dk_{2\,x}|\Psi(k_{1\,x},k_{2\,x},0,0)|^2.
\end{equation}
At the same values of all parameters which were used in the previous section, Eq. (\ref{plane}) would give the curve for the single-particle distribution shown in Fig. \ref{Fig8}, which differs drastically from that of Fig. \ref{Fig5}.
Such distributions are known \cite{scr} but, as shown above, the use of all photons of the emission cone gives a much wider information about features of noncollinear biphoton states which appear to have a much higher entanglement resource than it could be found from theoretical or experimental investigations of only photons propagating in a single given plane (e.g. $(xz)$).

\section{Conclusion}
To summarize, analysis of the noncollinear SPDC in the representation of Cartesian components of biphoton wave vectors shows that such states possess a very high resource of entanglement. As shown, this high entanglement is related to the described unusual and earlier not known features of the single-particle distribution of photons in the projection of their wave vectors on any given direction (e.g., $k_{x}\| 0x$). The shape of this distribution shown in Figs. $3,\,5$ is characterize by two well separated peaks at maximal values of $|k_{x}|$ and a long plateau-type region between them. Such shape differs drastically from the Gaussian-like curves, one example of which is given by the curve $(c)$ of Fig. \ref{Fig6}.  As shown, a very large broadening of this distribution is related to the noncollinearity of the SPDC process itself, and broadening owing to this mechanism can exceed significantly the usually described broadening related to the finite length of a crystal where SPDF emission is formed [see the analysis around Fig. \ref{Fig6} and the validity condition of Eq. (\ref{condition})]. A scheme of experiment in which such distribution can be seen is suggested and discussed. Roughly, to see the described effects one has to take into account contributions of all photons from the emission cone rather than only photons propagating in any single given plane. The derived results are in complete agreement with the earlier prediction of the enormous azimuthal entanglement \cite{PRA-16}, analysis of which was based on the consideration in the spherical-angle representation of transverse wave vectors of photons. But manifestation of this very high entanglement in spherical-angle and Cartesian pictures are absolutely different. Also, the suggested schemes for experimental observation of the described effects are different in this work and in \cite{PRA-16}. No doubts, realization of such experiments would be very interesting and, possibly, fruitful for application.

\section*{Acknowledgement}
The work is supported by the Russian Science Foundation, grant 14-02-01338-$\Pi$

\bibliography{Cart2}
\end{document}